\documentclass[reprint,amsmath,amssymb,aps,prx]{revtex4-2}

\usepackage{graphicx}% Include figure files
\usepackage{color}
\usepackage{comment}
\usepackage{multirow}
\usepackage{bm}% bold math
\usepackage[breaklinks=true,colorlinks,citecolor=blue,linkcolor=blue,urlcolor=blue]{hyperref}

\def\be {\begin{equation}}
\def\ee {\end{equation}}
\def\bea{\begin{eqnarray}}
\def\eea{\end{eqnarray}}

\begin{document}

%\title{Unidirectional magnetoresistance and spin-orbit torque in ${\mathcal P}{\mathcal T}$-symmetric 2D antiferromagnet CrSBr}
%\title{Nonlinear transport and magnetoelectric responses in ${\mathcal P}{\mathcal T}$-symmetric 2D magnet CrSBr}
%\title{Nodal line topology determines the quantum geometry induced transport and spin responses in CrSBr}
\title{Surface-Dominated Quantum Metric-Induced
Nonlinear Transport in the Layered
Antiferromagnet CrSBr}
\author{Kamal Das}
\affiliation{Department of Condensed Matter Physics, Weizmann Institute of Science, Rehovot 7610001, Israel}
\author{Yufei Zhao}
\affiliation{Department of Condensed Matter Physics, Weizmann Institute of Science, Rehovot 7610001, Israel}
\author{Binghai Yan}
\email{binghai.yan@weizmann.ac.il;  binghai.yan@psu.edu}
\affiliation{Department of Condensed Matter Physics, Weizmann Institute of Science, Rehovot 7610001, Israel
\\
Department of Physics, The Pennsylvania State University, University Park, Pennsylvania 16802, USA}

\begin{abstract}
The van der Waals (vdW) antiferromagnet CrSBr has recently garnered significant attention due to its air stability, high magnetic transition temperature, and semiconducting properties. We investigate its nonlinear transport properties and identify a quantum metric dipole-induced nonlinear anomalous Hall effect and nonlinear longitudinal resistivity, which switch sign upon reversing the N\'eel vector. The significant quantum metric dipole originates from Dirac nodal lines near the conduction band edge within the experimentally achievable doping range. Known the weak interlayer coupling, it is unexpected that the nonlinear conductivities do not scale
with sample thickness but are dominantly contributed by surface layers. In the electron-doped region, the top layer dominates the response while the top three layers contribute the most in the hole-doped region. Our results {establish topological nodal lines as a guiding principle to design high-performance nonlinear quantum materials and} suggest that surface-sensitive transport devices will provide new avenues for nonlinear electronic applications.

\end{abstract}

\maketitle

%\section{Introduction}

%%%% Setting the context of the paper, why CrSBr is interesting %%%%%%
Magnetism in two-dimensional (2D) vdW materials has recently gained intensive attention, challenging the long-held belief that intrinsic magnetic order could not be sustained in 2D due to enhanced thermal fluctuations~\cite{burch_N2018_magnetism,wang_ACSN2022_the,ahn_PQE2024_progress}. Although several 2D ferromagnets (FMs) and antiferromagnets (AFMs) have been demonstrated~\cite{huang_N2017_layer,gong_N2017_discovery,lee_NL2016_ising,ni_NN2021_imaging,li_SA2019_intrinsic,rahman_ACSN2021_recent,olsen_2DMat2024_anti,shao2025magnetically}, most suffer from low magnetic order temperatures or extreme air sensitivity, limiting their further studies and practical applications. In this context, CrSBr has emerged as a promising vdW AFM with a simple magnetic structure characterized by spins along the $b$-axis (AFM-$b$), exhibiting a high Néel temperature of 132 K, substantial air stability, and semiconducting properties~\cite{goser_JMMM1990_magnetization,telford_AM2020_layered,telford_NM2022_coupling,lopez-paz_NC2022_dynamic,lee_NL2021_magnet,wilson_NM2021_inter,liu_ACSN2022_a,rizzo_AM2022_visual,dirnberger_N2023_magneto,guo_NC2024_extraordinary, vakili_NC2024_doping,liu_arxiv2024_probing,bagani_arxiv_2024imaging}. Unlike most vdW magnets, CrSBr displays in-plane magnetization~\cite{telford_AM2020_layered,telford_NM2022_coupling,lopez-paz_NC2022_dynamic}, strong charge-spin \cite{telford_AM2020_layered,telford_NM2022_coupling} and magneto-optical coupling~\cite{wilson_NM2021_inter,dirnberger_N2023_magneto,vakili_NC2024_doping}, quasi-1D behavior~\cite{wu_AM2022_quasi,klein_ACSN2023_the}, and controllable magnetic ordering~\cite{telford_APR2023_design,wang_PRB2023_magnetic, cenker_NN2022_rever}. Its stability down to a few layers~\cite{lee_NL2021_magnet,telford_NM2022_coupling} enables enhanced control over its properties, making CrSBr a promising candidate for future spintronic applications~\cite{jungwirth_NN2016_anti,baltz_RMP2018_anti,gish_NE2024_vdw}.

The layered AFM structure of CrSBr is similar to the known AFM topological insulator MnBi$_2$Te$_4$ which exhibits notable nonlinear transport phenomena in thin films ~\cite{wang_N2023_quantum,gao_SC2023_quantum}. While nonlinear responses have predominantly been investigated in topological materials due to their inherent quantum geometric effects \cite{nagaosa_ARCMP2024_nonreci,ideue_ARCMP2021_symmetry,jiang_arxiv2025_revealing}, we point out that generic topological features near the Fermi energy, such as band crossing or anti-crossing, will generate substantial Berry curvature or quantum metric even in an ordinary material. Therefore, we are motivated to explore topological characteristics in the band structure of CrSBr and study possible nonlinear transport phenomena when the chemical potential is tuned close to the topological region by doping.

%%%%%%%%%%%%%%%%%%%%%%%%%%%%%%%%%%%%%%%%%%%%%%%%%%%%%%%%%
\begin{figure*}[t]
\centering
\includegraphics[width =  \linewidth]{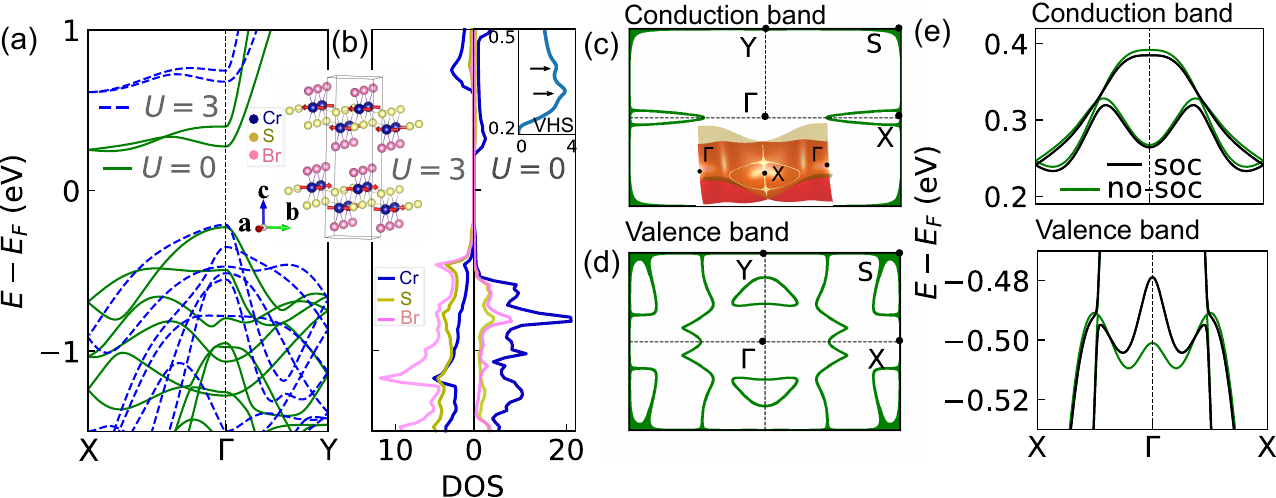}
\caption{\textbf{Crystal structure, electronic bands, and nodal lines of bulk CrSBr.} (a) Band dispersion from GGA calculations without (solid green) and with (blue dashed) Hubbard $U$ in the absence of spin-orbit coupling. The inset shows the AFM-$b$ phase of bulk CrSBr. (b) Orbital projected density of states (DOS) for $U=3$ eV (left panel) and $U=0$ eV (right panel). The inset highlights the total DOS near the conduction band minima for $U=0$ eV, which shows the van-hove singularity (VHS). (c)-(d) Distribution of the Dirac nodal lines in the $k_x-k_y$ plane for the first conduction and valence band, respectively.  (e) The opening of the band gap in the presence of SOC in the conduction (top) and valence (bottom) band.
}
\label{fig_1}
\end{figure*}
%%%%%%%%%%%%%%%%%%%%%%%%%%%%%%%%%%%%%%%%%%%%%%%%%

%%%% what do we do in this paper? %%%%%%%%
Remarkably, the band structure of CrSBr exhibits Dirac nodal lines within the experimentally accessible energy range, at about {$63$ meV} above the conduction band minima (CBM) and 250 meV below the valence band maxima (VBM). These nodal lines generate a significantly large quantum geometry and lead to strong nonlinear transport. The even number of layers breaks inversion symmetry ($\mathcal{P}$) but preserves the spatial-time reversal symmetry $({\mathcal P} {\mathcal T})$, displaying the quantum metric dipole-induced nonlinear anomalous Hall effect (NLAHE) and nonlinear longitudinal resistance -- an effect observed only in a few magnetic materials so far~\cite{kaplan_PRL2024_uni,wang_N2023_quantum,gao_SC2023_quantum,watanabe_JPCM2024_magnetic}. Intriguingly, the nonlinear conductivities in thick films does not scale with the sample thickness but is dominated by the surface layers, different from the linear transport, because surface layers exhibit strong inversion-breaking while the inner layers do not. 
The outermost layer primarily dominates the nonlinear response in the electron-doped region, and the three outermost layers dominate in the hole-doped region. Our findings demonstrate topological nodal line as a recipe of large nonlinear responses and highlight that the transport properties of CrSBr can be effectively probed using surface-sensitive techniques, for example, by placing the contact on the surface.

%%% Debate of electronic structure; why U=0 is the best on DFT level?
We first introduce the band structure. Recent studies reported distinct band structures with different methods~\cite{klein_ACSN2023_the,guo_SS2024_second,watson_npj2D2024_giant}. Transport and optical measurements~\cite{wu_AM2022_quasi,klein_ACSN2023_the} showed a quasi-1D nature of material properties but varied band gaps. Recent angle-resolved photoemission spectroscopy (ARPES) data in paramagnetic~\cite{bianchi_PRB2023_paramag,smolenski_arxiv2024_large} and AFM phase~\cite{bianchi_PRB2023_chrage,watson_npj2D2024_giant,wu_arxiv2024_mott} revealed more information of the band structure. Studies~\cite{watson_npj2D2024_giant,wu_arxiv2024_mott} reveal that the conduction bands exhibit a quasi-1D character, while the valence bands exhibit a quasi-2D nature. Top valence bands are contributed by the Cr$-t_{2g}$ orbital while the lower valence bands are predominated by Br/S$-p$ orbitals. Several valence bands located closely in energy at the top of $X$ point. These ARPES features was reproduced by GW and dynamical mean field theory calculations~\cite{watson_npj2D2024_giant,wu_arxiv2024_mott}. Although it produces a small band gap, similar can be captured on the generalized gradient approximation (GGA) of the density functional theory (DFT) using Hubbard $U=0$ eV. 
We show the band structure of the bulk CrSBr in the AFM-$b$ phase for $U=0$ eV in Fig.~\ref{fig_1}(a). 
In contrast, GGA+U calculations show different features from ARPES, although it is widely used in literature. 
For example, the GGA+U top valence bands at $X$ are sparsely spaced and Cr-$t_{2g}$ orbitals are over-hybridized with other orbitals as shown in Fig.~\ref{fig_1}(b). In addition, we find that DFT ($U=0$ eV) correctly reproduces the AFM phase as the ground state while DFT+U ($U=3$ eV) shows the FM phase as the ground state in calculations, consistent with a recent work~\cite{wang_PRB2023_magnetic}.  For these reasons, we will adopt GGA ($U=0$ eV) calculations in this study. See details {in S1 and S2 of} Supporting Information (hereafter SI).

%%% Discussing the band crossing and nodal lines; why previous literature missed it 
We find a band crossing along the $\Gamma-X$ path near the CBM while the valence band shows multiple crossings. The crossing points in the conduction bands extend to form a $X$-centric Dirac nodal ring as shown in Fig.~\ref{fig_1}(c). In the valence band, the crossings extend into a more complex structure of Dirac nodal lines. Figure~\ref{fig_1}(d) shows the nodal lines for the first valence band. These Dirac-type band crossings are protected by the glide mirror symmetry ${\overline M}_z \equiv \{ {M_z}| \tau \}$, where $\tau=(\frac{1}{2},\frac{1}{2},0)$ is the translation, and the nodal lines lie within the glide mirror plane ($k_z=0$). We mention that the precise location of the crossing and the resultant nodal lines are sensitive to the atomic positions and the correlation parameter $U$. For example, some studies report crossing near the CBM along the $\Gamma-{ Y}$ path~\cite{yang_PRB2021_triaxial,bo_NJP2023_calculated,klein_ACSN2023_the,bianchi_PRB2023_chrage}, which is likely based on a relaxed lattice structure without including magnetism (See {S2 of } SI for details). Some calculations with $U=3 $ eV do not show any crossing near the CBM ~\cite{guo_NS2018_chromium,wang_APL2020_elec,guo_SS2024_second} while the band crossing along $\Gamma-{X}$ is consistent with Refs.~\cite{wilson_NM2021_inter,wang_PRB2023_magnetic,datta_arxiv2024_magnon}. Future ARPES studies are called for the confirmation of these nodal lines. 

%%%%%%%%%%%%%%%%%%%%%%%%%%%%%%%%%%%%%%%%%%%%%%%%%%
\begin{figure*}[t]
\centering
\includegraphics[width =  \linewidth]{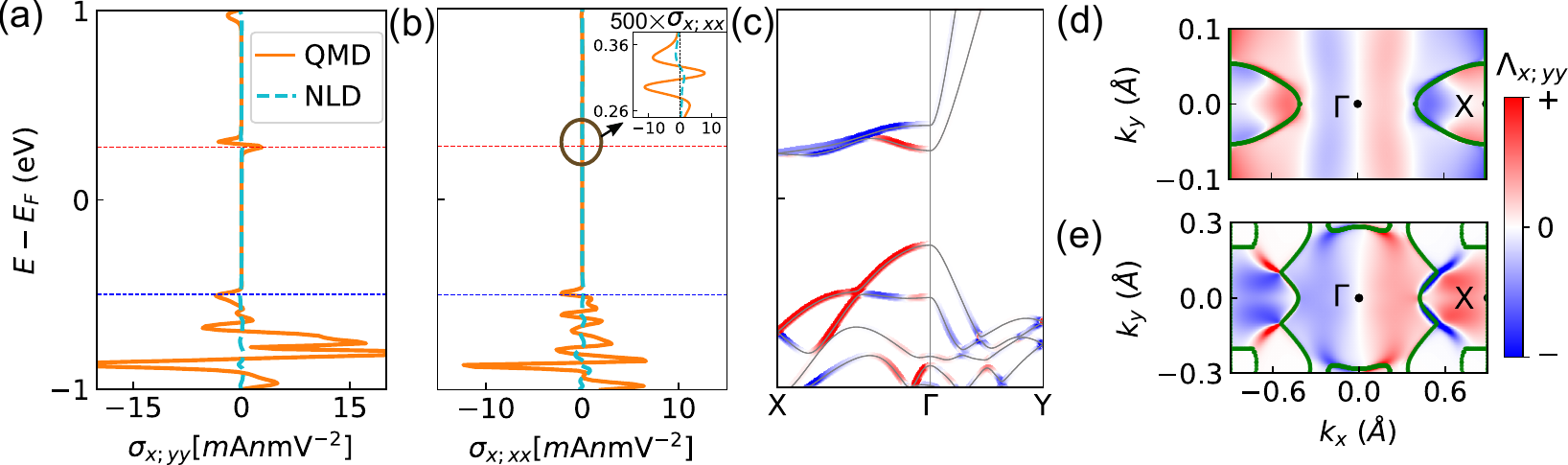}
\caption{{\bf Nonlinear charge responses in bilayer CrSBr:}  (a) The nonlinear transverse conductivity  $\sigma_{x;yy}$ and (b) longitudinal conductivity $\sigma_{x;xx}$ for bilayer CrSBr in AFM-$b$ phase. The orange solid and the cyan dashed lines are the contributions from the quantum metric dipole and nonlinear Drude, respectively. The inset in (b) shows contribution near the conduction band minima. (c) The band dispersion weighted by quantum metric dipole for the $\sigma_{x;yy}$ conductivity. (d)-(e) The momentum space distribution of the quantum metric dipole for the first conduction and valence bands. The high-concentration region follows the nodal line shown by the green lines.  }
\label{fig_2}
\end{figure*}
%%%%%%%%%%%%%%%%%%%%%%%%%%%%%%%%%%%%%%%%%%%%%%%%%%%

%%% what happens to Nodal lines the monolayer and bilayer
This nodal line topology of the bulk is inherited from the monolayer symmetry. The monolayer, and odd number of layers share the same glide mirror with the bulk where the mirror plane lies inside the vdW layer {and can be represented by the spin group $\{ {\mathcal I}  || \overline M_{1z} $\}, ${\mathcal I}$ being the identity operation on spin}. Therefore, they present the same Dirac crossing in the absence of spin-orbit coupling (SOC). In an even number of layers, the glide mirror plane shifts to the vdW gap and the glide mirror {$\overline M_{2z}$ connects the opposite spins of the adjacent layers. Consequently, it involves an extra spin rotation with the spin group operation $\{ {{\mathcal U}(\pi)} || \overline M_{2z}  $\}.}  Thus, the Dirac nodal lines are weakly gapped due to symmetry reduction even without SOC in the even number of layers. After including SOC, all these nodal lines are weakly gapped, regardless of layer thickness {(See S3 of SI for more details)}. Such gap opening for the bulk is highlighted in Fig.~\ref{fig_1}(e) in the conduction (top panel) and valence (bottom panel) band along the $X-\Gamma-X$ path. It is known that the gapped Dirac crossing (or anti-crossing) commonly contributes a significant Berry curvature and leads to large anomalous Hall effect~\cite{Li2020giant} or spin Hall effect~\cite{Sun2017dirac} in nodal line semimetals. %{As we will see below, the gapped Dirac crossings in CrSBr also generate significant quantum metric~\cite{feng_arxiv2024_quantum} which primarily determines the nonlinear transport properties.} 
As we will see below, the gapped Dirac crossings in CrSBr also {generate large quantum metric~\cite{feng_arxiv2024_quantum} which significantly enhances the nonlinear transport properties.} Importantly, the nodal lines lie within the accessible energy range, in particular on the conduction band side. In an early experiment~\cite{telford_AM2020_layered}, for example, CrSBr is electron-doped to 2D carrier density $n_{2D}\approx5\times10^{13}$ cm$^{-2}$, corresponding to the chemical potential $\approx 70$ meV above CBM, which is close to the Dirac nodal line. Furthermore, in the charge-neutral pristine samples doping of alkali atoms (\textit{e.g.} Li, K) can effectively shift the chemical potential as seen in early experiments~\cite{bianchi_PRB2023_chrage,smolenski_arxiv2024_large,Feuer2025}. 

%%%%%%%%%%%%%%%%%%%%%%%%%%%%%%%%%%%%%%%%%%%%%%%%%%%%%%%%%%%%%%%%%%%
\begin{table}[b]
\caption{The magnetic point group (MPG) for even and odd-layer CrSBr for the three spin orientations in the AFM and FM phases. Listed the allowed nonlinear transport (${J}_{i}=\sigma_{i;jl}E_j E_l $ with $\sigma_{i;jl}=\sigma_{i;lj}$), and linear anomalous Hall (${J}_{i}=\sigma_{ij}^{\rm AHE}E_j $ with $\sigma_{ij}^{\rm AHE}=-\sigma_{ji}^{\rm AHE}$) coefficients where $i,j,l \in x,y $. The key symmetries that forbid the response coefficients are highlighted. }
\label{sym_table}
\setlength{\tabcolsep}{1pt} % Default value: 6pt
\renewcommand{\arraystretch}{1.5} % Default value: 1
\begin{tabular}{c c c c c c }
 &   & {\bf Even layers} &  &  \\
\hline \hline 
Spin axis & MPG  & Key symmetry & $\sigma^{\rm AHE}_{ij}$ & $\sigma_{i;jl}$ \\
\hline \hline
AFM-$b$ &  $m'mm$ & $\overline M_y$ & $\times$  & $\sigma_{x;xx};\sigma_{x;yy};\sigma_{y;xy}$  
\\
AFM-$a$  &  $mm'm$   & $\overline M_x$   & $\times$ & $\sigma_{y;xx};\sigma_{y;yy};  \sigma_{x;xy}$
\\
AFM-$c$  & $m'm'm'$  & $\overline M_y {\mathcal T}$, $\overline M_x {\mathcal T}$, ${\mathcal P} {\mathcal T}$ & $\times$ & $\times$  \\
FM-$b$ & $m'mm'$  & $\overline M_y$, ${\mathcal P}$  &$\times$  & $\times$\\ 
FM-$a$ & $mm'm'$  & $\overline M_x$, ${\mathcal P}$ &$\times$  & $\times$\\ 
FM-$c$ & $m'm'm$  &  ${\mathcal P}$ &$\sigma_{xy}^{\rm AHE}$  & $\times$ \\
\hline 
 &   & {\bf Odd layers} &  &  
\\
\hline 
AFM/FM-$b$ & $m'mm'$ & $\overline M_y$, ${\mathcal P}$ & $\times$ &$\times$  \\
AFM/FM-$a$ & $mm'm'$  & $\overline M_x$, ${\mathcal P}$  & $\times$ & $\times$ 
\\
AFM/FM-$c$ & $m'm'm$  &  ${\mathcal P}$  & $\sigma_{xy}^{\rm AHE}$  & $\times $  \\ 
\hline \hline
\end{tabular}
\end{table}
%%%%%%%%%%%%%%%%%%%%%%%%%%%%%%%%%%%%%%%%%%%%%%%%%%%%%%%%%%%%%%%%%%%%%%

%%%%%%%%%%%%%%%%%%%%%%%%%%%%%%%%%%%%%%%%%%%%%%%%%%
\begin{figure*}[t!]
\centering
\includegraphics[width = \linewidth]{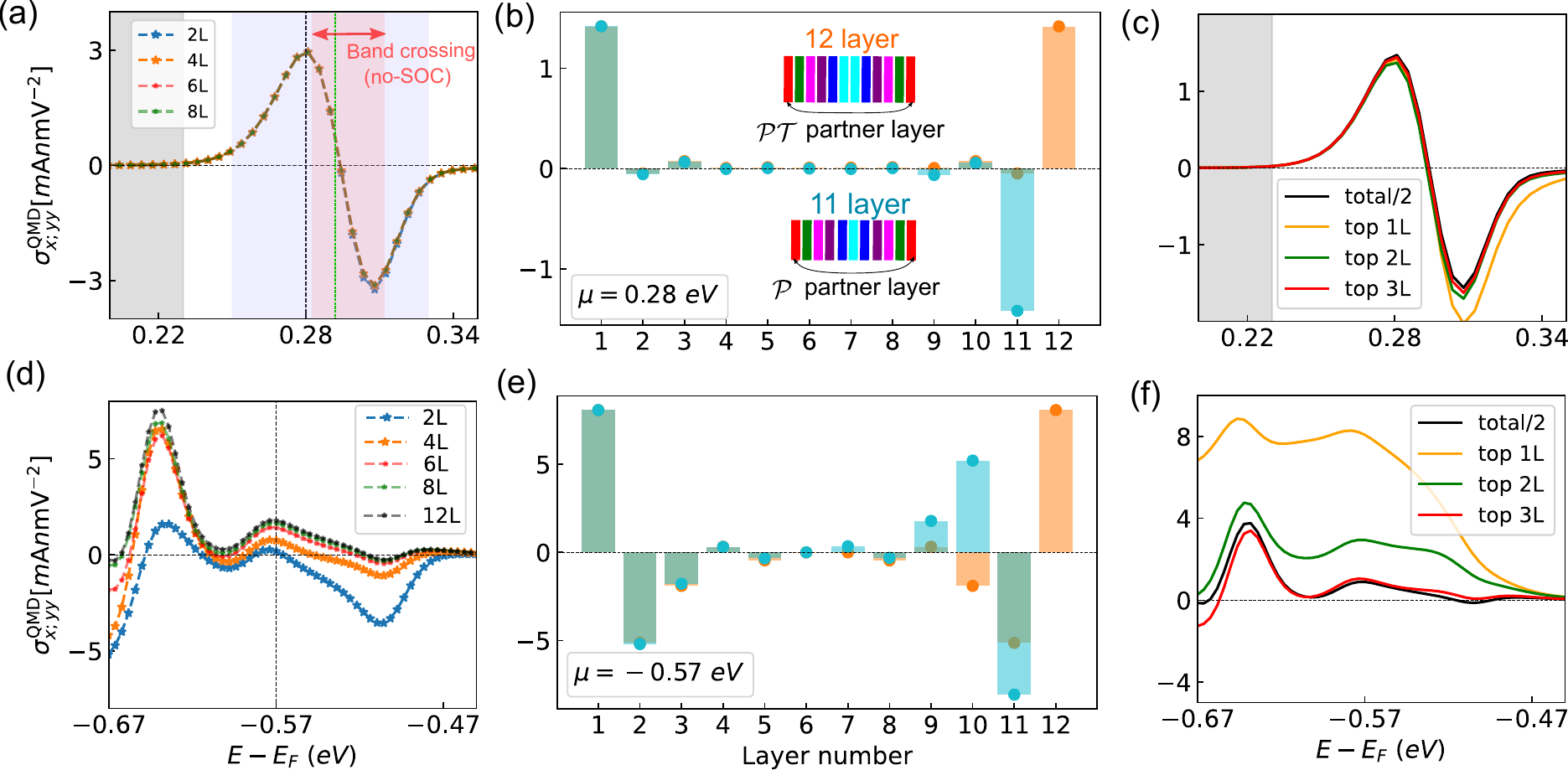}
\caption{{\bf Surface dominated nonlinear transport in thick films:} The variation of the quantum metric dipole contribution with chemical potential for films with different thickness in the conduction band sector (a) and valence band sector (d). The grey shaded region shows the band gap. { The red and blue shaded regions in (a) show the nodal line energy range and significant conductivity region. The green dashed line show the experimental chemical potential~\cite{telford_AM2020_layered}.} The layer projection of the nonlinear conductivity for 12L (orange) and 11L (cyan) thick films at $\mu=0.28$ eV (b) and at $\mu=-0.57$ eV (e). In the schematic of slabs, the same colored layers are a parity-time reversal partner in 12L film and inversion partners in 11L. The nonlinear conductivity from 1L, 2L, and 3L layers near the surface is compared with half of the total contribution (black) in the conduction band (c) and valence band (f).}
\label{fig_3}
\end{figure*}
%%%%%%%%%%%%%%%%%%%%%%%%%%%%%%%%%%%%%%%%%%%%%%%%%%%

%%%%%%%%%% general introduction to nonlinear transport and why qmd is important %%%%%%%%%
With a comprehensive understanding of the band structure in place, we now move on to the nonlinear response of even-layer films of CrSBr. {The response tensor is qualitatively determined by the space-time symmetries.} Due to the combined ${\mathcal P} {\mathcal T}$ symmetry, the Berry curvature dipole~\cite{sodemann_PRL2015_quant} vanishes in this system. Hence, the nonlinear conductivities (NLCs) within the relaxation time approximation~\cite{gao_PRL2014_field,das_PRBL2023_intrin,kaplan_PRL2024_uni} are given by,
\begin{subequations} 
\begin{align} \label{NLD}
\sigma_{i;jl}^{\rm NLD} &= -\tau^2\dfrac{e^3}{ \hbar^3}\sum_n\int_{\bm k} f_n \partial_i \partial_j \partial_l  \epsilon_{n} ~, 
\\   
\sigma_{i;jl}^{\rm QMD}  &= - \dfrac{e^3}{\hbar} \sum_{n} \int_{\bm k} f_n \big[ 2 \partial_i   {\mathcal G}^n_{jl} - \frac{1}{2} \left( \partial_{j}  {\mathcal G}^n_{il} + \partial_{l} {\mathcal G}^n_{ij}\right) \big]. \label{QMD}
\end{align}
\end{subequations}
Here, $\partial_i$ stands for $\partial/\partial {k_i}$, $f_n$ for the Fermi function, $\epsilon_n$ for the energy, and ${\mathcal G}_{ij}^n$ for the band normalized {intra-band} quantum metric given by ${\mathcal G}_{ij}^{n} =\sum_{p\neq n}  2 {\rm Re} [\langle u_n  |\partial_i {\mathcal H}| u_p \rangle \langle u_p  |\partial_j {\mathcal H} | u_n  \rangle]/(\epsilon_n -\epsilon_p)^3$. Equation~\eqref{NLD} represents the nonlinear Drude (NLD) contribution which originates from the current-induced Fermi surface's shift and is completely determined by the dispersive velocities of the wave packet. Contrary to this, Eq.~\eqref{QMD}  originates from the quantum metric dipole (QMD) defined as
\begin{equation} \label{qmd}
\Lambda_{i;jl}^n = 2 \partial_i   {\mathcal G}^n_{jl} - \frac{1}{2} \left( \partial_{j}  {\mathcal G}^n_{il} + \partial_{l} {\mathcal G}^n_{ij} \right)~,
\end{equation}
which is determined by the first derivative of the band-normalized quantum metric. Equation~\eqref{QMD} has drawn attention recently because, unlike the Berry curvature, the quantum metric {that represents the distance between two neighboring states in the momentum space} hardly appears in transport, and this contribution is scattering time independent, being completely determined by the band structure (intrinsic). Furthermore, Eq.~\eqref{QMD} has both longitudinal and transverse contributions. The longitudinal part causes a quantum geometric contribution to the unidirectional resistance, while the transverse part gives rise to an intrinsic NLAHE~\cite{gao_SC2023_quantum,han_NP2024_room}.

{In the ground state AFM-$b$, even-layer films belong to the magnetic point group $m^\prime m m$ (magnetic space group $Pm^\prime m n$) which includes the symmetries $\overline M_y\equiv \{M_y | (0,\frac{1}{2},0)\}$, $\overline M_x {\mathcal T}\equiv \{M_x | (\frac{1}{2},0,0) \} {\mathcal T}$, and $\overline  M_z$. The in-plane nonlinear conductivities are determined by $\overline M_y$ and $\overline M_x {\mathcal T}$ symmetries, which impose same constraints on the response tensors. Specifically, these symmetries forbid conductivities with odd number of $y$-indices: $\sigma_{y;yy}, \sigma_{y;xx}$ and $\sigma_{x;yx}(=\sigma_{x;xy})$. This can be understood by analyzing the real space transformation of current and electric fields under $\overline M_y$ where current transforms as $j_y \to -j_y$, $j_x \to j_x$ and the electric field as $E_y \to -E_y$ and $E_x \to E_x$. Consequently, only conductivities with even number of $y$, namely $\sigma_{x;xx}, \sigma_{x;yy}$ and $\sigma_{y;yx}=\sigma_{y;xy}$ are allowed. Additionally, the mirror symmetry $\overline M_y$ forbids linear anomalous Hall conductivity. The response tensors for various spin orientations in the FM and AFM phases are highlighted in Table.~\ref{sym_table} (See S6 of SI for details).} We present the variation of NLC $\sigma_{x;yy}$ and $\sigma_{x;xx}$ with the chemical potential for the bilayer system in the ground state in Figs.~\ref{fig_2}(a) and (b), respectively. For scattering time $\tau=0.01$ ps, we find that the QMD dominates the NLD contribution. {Since the NLD contribution is more than an order smaller than the QMD conductivity, we will not discuss it further here, but will provide a detailed analysis in S6 of SI.} Interestingly, the conductivities peak only in a small energy window in the conduction band sector near $\mu=0.28$ eV (red dashed line). In the valence band sector, NLC becomes significant well inside the VBM after $\mu =-0.5$ eV (blue dashed line). Notably, $\sigma_{x;xx}$ is of the same order of magnitude compared to $\sigma_{x;yy}$ on the valence band side but much smaller on the conduction band side, see the inset of Fig.~\ref{fig_2}(b). This is due to the quasi-1D and quasi-2D nature of the conduction and valence bands, respectively. {We mention that the NLCs are calculated for a specific N\'eel configuration. When the spin orientations are reversed {\it i.e.} up-down becomes down-up or vice versa, the NLCs changes sign~\cite{liu_PRL2021_intrinsic}. 

%%% This paragraph highlights the role of nodal line topology %%%%%%%%%%
The nodal lines play a pivotal role in shaping the NLCs. Specifically, peaks and sign changes near $\mu=0.25-0.35$ eV and the onset of a significant response after $\mu=-0.5$ eV are directly related to the position of the nodal lines, as demonstrated below for $\sigma_{x;yy}^{\rm QMD}$. The QMD weighted band dispersion in Fig.~\ref{fig_2}(c) shows that all regions in the chemical potential where there is a peak in NLC correspond to some band crossing and the resultant large QMD. This behavior is further elucidated by the momentum-resolved QMD in the 2D Brillouin zone. The high intensity of QMD in Fig.~\ref{fig_2}(d), which roughly follows the nodal line structure shown by green line, causes the peak in the NLC in between $\mu=0.28-0.35$ eV. In the valence band sector also, the large QMD follows the nodal line as shown in Fig.~\ref{fig_2}(e). Relatively large contribution appears after $\mu=-0.5$ eV compared to $\mu=-0.4$ eV as the Fermi contour intersects the nodal lines with a high density of QMD as illustrated in Fig. S3(f) of SI. This behavior is also associated with the orbital weight of magnetic Cr atoms, which is small for the valence band top but increases significantly below $\mu = -0.5$ eV. Because magnetic states primarily govern the NLCs, the contribution remains small near the VBM. This makes the detection of topological nodal line and the resulting nonlinear transport more feasible to measure on the conduction band side. In Fig.~\ref{fig_3}(a) we have depicted the energy range of the nodal line (63-87 meV, red shaded region) and region of significant response (blue shaded) along with the experimentally reported chemical potential dashed green~\cite{telford_AM2020_layered} lines. Notably, the region of significant response extends beyond the nodal line region as SOC opens a slight gap, and consequently, the QMD spreads below and above the gap. The envelope-like profile of the nonlinear response centered around the nodal line region is a characteristic feature of quantum geometry-driven properties in both gapped and gapless systems, reflecting their underlying topological origin.

Optical and transport properties shows a strong 2D nature with weakly-coupled layers for CrSBr in optical and transport measurements~\cite{klein_ACSN2023_the,guo_NC2024_extraordinary,lin_PRR2024_influence,scheie_AS2022_spin}. However, NLCs distribute nonuniformly among layers in a thin film. 
Here, we resolve the layer by layer contribution to NLC by using thick films as examples.  In Fig.~\ref{fig_4}(a), we show the QMD-induced NLC from 2L-8L in the conduction band sector near $\mu = 0.25-0.35$ eV. Intriguingly, the NLC does not scale with layer number and is almost the same for all. This is related to the fact that the surface layer has a dominant contribution to NLC. Furthermore, we plot the layer projection of the QMD-induced NLC at $\mu=0.28$ eV for 12L- and 11L-thick films in Fig.~\ref{fig_4}(b). For both films, the inner layer contributes little and most of the contribution arises from the outermost layer. Notably, the layers from the top and bottom sides have the exactly same contributions for the 12L-thick film, as they are ${\mathcal P} {\mathcal T}$ partners of each other. For the 11L-thick film, the outer layers have opposite contributions as they are connected by ${\mathcal P}$ symmetry, leading to a vanishing total response. %We note that the outermost layers have identical contributions for the even and odd number of layers. 
In Fig.~\ref{fig_4}(c) we compared half of the total contribution (black line) of the 12L film with the contribution from the top 1L (orange), 2L (green), and 3L (red). It is evident that the surface layers completely determine the total response of the thick film.

%%%%%%%%%%%%%%%%%%%%%%%%%%%%%%%%%%%%%%%%%%%%%%%%%%
\begin{figure}[t!]
\centering
\includegraphics[width = 1\linewidth]{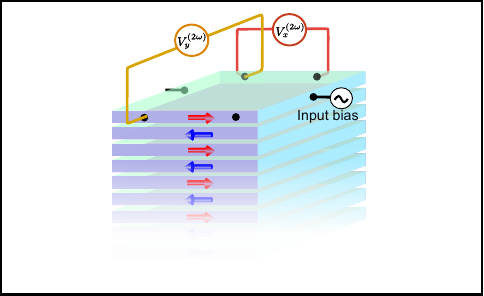}
\caption{{\bf Schematic of surface sensitive nonlinear electronic device:} Nonlinear transport through the source and drain grown on the surface. The yellow channel represents Hall and red channel represents longitudinal nonlinear transport. %{for the NLHE, there should be only one transverse current direction preferred}
}
\label{fig_4}
\end{figure}
%%%%%%%%%%%%%%%%%%%%%%%%%%%%%%%%%%%%%%%%%%%%%%%%%%%

A similar layer projection analysis for the valence bands is shown in Fig.~\ref{fig_4}(d)-(f). The NLC for 2L-8L and 12L is presented in Fig.~\ref{fig_4}(d). As observed in the conduction bands, the NLC does not scale with the number of layers. However, unlike the conduction bands, the valence band response requires more layers to reach saturation, which occurs after 8L. 
%This suggests that the system attains bulk-like behavior beyond 8L , consistent with recent observations of the bulk Néel temperature for this thickness ~\cite{watson_npj2D2024_giant}. ===== bulk has no NLC, this is misleading 
The higher number of layers required for saturation in the valence bands can be attributed to the stronger inter-layer coupling through the atomic orbital dominant in these bands. The interlayer coupling occurs primarily through Br-$p$-orbitals~\cite{wang_PRB2023_magnetic}, as the Cr-S layers are sandwiched between Br layers. The Br-$p$ orbitals contribute significantly in the valence band top while 
their contribution at the conduction band bottom is vanishingly small. Consequently,
interlayer coupling is much weaker in these conduction bands than valence bands.
The layer projected contribution at $\mu =-0.57$ eV is shown in Fig.~\ref{fig_4}(e). Here, we note that only the outermost layer is insufficient to describe the overall transport and one needs to consider the outer three layers. The corresponding chemical potential dependence of the outermost layer's contribution is highlighted in Fig.~\ref{fig_4}(f) implying that the outermost three layers determine the total response.

%%%%%%%%%%%%%%%%%%%%%%%%%%%%%%%%%%%%%%%%%%%%%%%
%\section{Discussion}
%\label{Discussions]

The reason for the surface layer domination can be attributed to symmetry. Each individual layer of CrSBr has inversion symmetry. Due to this, layers that are deep inside the slab do not experience inversion breaking. However, the outermost layers feel the inversion breaking due to the surface termination and hence make dominant contribution to the nonlinear transport. We should point out that surface nonlinear Hall effects were reported for topological materials~\cite{he_NC2021_quantum, makushko_NE2024_tunable}, where topological surface states lead to Berry curvature-related extrinsic nonlinear responses. Different from them, the CrSBr film exhibit no surface state and every layer conducts electrons in ordinary transport. Therefore, it becomes nontrivial that the surface layer becomes dominant in the nonlinear transport. In addition, 
we note that a similar surface contribution to the nonlinear optical effect was proposed in another layered antiferromagnet CrI$_3$~\cite{zhou_PRL2024_skin}, in which the inter-band quantum metric {representing the dipole transition between bands} was discussed. Our studies focus on the topological nodal line induced intrinsic nonlinear transport, despite that extrinsic effects may also exist in experiments~\cite{huang_arxiv2023_scaling,gong_arxiv2024_nonlinear,mehraeen_PRB2024_quantum}.

%%%%%%%%%%%%%%%%%%%%%%%%%%%%%%%%%%%%%%%%%%%%%%%
%\section{Conclusion}
%\label{conclusion

In conclusions, the nonlinear transport is predominantly surface-driven, insensitive to the film thickness in CrSBr. 
This behavior arises from the surface inversion-breaking and local single layer inversion symmetry, which is even independent of the underlying magnetic order (\textit{e.g.}, AFM or FM). Our findings go beyond the expected surface transport of axion insulators or topological insulators, in which topological surface states constitute the Fermi surface. 
Our results indicates that surface-sensitive measurements, where contacts are placed on a single surface of a film as shown in Fig.~\ref{fig_4}, can effectively probe the nonlinear responses and quantum geometry. The sample thickness can be less sensitive in the device fabrication compared to ordinary thin film devices~\cite{gao_SC2023_quantum,wang_N2023_quantum}.
Our calculations indicate that the electron-doped sample will exhibit much stronger nonlinear responses than the hole-doped case. {While CrSBr serves as the material platform in our study, the core findings---the surface dominant quantum geometry and its origin in topological band structure provide a general guideline and are broadly applicable to other materials. Our work paves a path to design surface-sensitive nonlinear devices for applications in energy-harvesting, photodetectors, and spintronics. 

{\it Note added: } During the review of our manuscript, we became aware of an experimental study~\cite{jo_AM2025_anomalous} which measure longitudinal nonlinear conductivity in CrSBr. They also concluded that QMD is the major contribution.

\section{Acknowledgement}
We thank Dr. Hengxin Tan for his help in developing Wannier function-related codes during the early stages and Prof. Yinming Shao for inspring discussions. 
K. D. acknowledges the financial support by the Weizmann Institute of Science, Dean of Faculty fellowship, and the Koshland
Foundation. B.Y. acknowledges the financial support by the Israel Science Foundation (ISF: 2932/21, 2974/23).

\bibliography{main}

\end{document}